\documentclass[format=sigconf, review=false, anonymous=False]{acmart}

\usepackage[subtle,leading=normal]{savetrees}

\usepackage{enumitem}
\usepackage{makecell}
\usepackage{float}
\usepackage{booktabs}
\usepackage{rotating}
\usepackage{extdash}
\usepackage{tabularx}
\usepackage{pifont}
\usepackage{tikz}
\usepackage{multirow}
\usepackage{hyphenat}
\usetikzlibrary{automata}
\usetikzlibrary{positioning}

\usepackage{graphicx}
\graphicspath{ {img/} }

\newcommand{\cybopt}{{\tt CySecTool}}

\title{Optimal Security Hardening over a Probabilistic Attack Graph}
\subtitle{A Case Study of an Industrial Control System using \cybopt{}}

\author{Przemysław Buczkowski}
\affiliation{%
\institution{Queen Mary University of London}
\department{School of Electronic Engineering and Computer Science}
\streetaddress{Mile End Road}
\city{London}
\state{England}
\postcode{E1 4NS}
\country{United Kingdom}}
\email{p.buczkowski@qmul.ac.uk}

\author{Pasquale Malacaria}
\affiliation{%
\institution{Queen Mary University of London}
\department{School of Electronic Engineering and Computer Science}
\streetaddress{Mile End Road}
\city{London}
\state{England}
\postcode{E1 4NS}
\country{United Kingdom}}
\email{p.malacaria@qmul.ac.uk}

\author{Chris Hankin}
\affiliation{%
\institution{Imperial College London}
\department{Institute for Security Science and Technology}
\streetaddress{180 Queen's Gate}
\city{London}
\state{England}
\postcode{SW7 2AZ}
\country{United Kingdom}}
\email{c.hankin@imperial.ac.uk}

\author{Andrew Fielder}
\affiliation{%
\institution{Imperial College London}
\department{Institute for Security Science and Technology}
\streetaddress{180 Queen's Gate}
\city{London}
\state{England}
\postcode{SW7 2AZ}
\country{United Kingdom}}
\email{a.fielder@imperial.ac.uk}

\copyrightyear{2022} 
\acmYear{2022} 
\setcopyright{rightsretained} 

\acmConference[SaT-CPS '22]{Proceedings of the 2022 ACM Workshop on Secure and Trustworthy Cyber-Physical Systems}{April 27, 2022}{Baltimore, MD, USA}
\acmBooktitle{Proceedings of the 2022 ACM Workshop on Secure and Trustworthy Cyber-Physical Systems (SaT-CPS '22), April 27, 2022, Baltimore, MD, USA}
\acmDOI{10.1145/3510547.3517919}
\acmISBN{978-1-4503-9229-7/22/04}

\usepackage{etoolbox}
\makeatletter
\patchcmd{\maketitle}{\@copyrightpermission}{
  \begin{minipage}{0.4\columnwidth}
    \href{http://creativecommons.org/licenses/by/4.0/}{\includegraphics[width=0.95\textwidth]{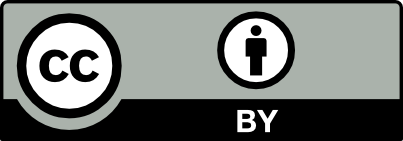}}
  \end{minipage}\hfill
  \begin{minipage}{0.6\columnwidth}
      \href{http://creativecommons.org/licenses/by/4.0/}{\nohyphens{This work is licensed under a Creative Commons Attribution International 4.0 License.}}
  \end{minipage}

  \vspace{5pt}
}{}{}

\makeatother

\begin{document}
\fancyhead{}
\begin{abstract}
\cybopt{} is a tool that finds a cost-optimal security controls portfolio in a given budget for a probabilistic attack graph. A portfolio is a set of counter-measures, or controls, against vulnerabilities adopted for a computer system, while an attack graph is a type of a threat scenario model. In an attack graph, nodes are privilege states of the attacker, edges are vulnerabilities escalating privileges, and controls reduce the probabilities of some vulnerabilities being exploited. The tool builds on an optimisation algorithm published by \cite{attackgraphs}, enabling a user to quickly create, edit, and incrementally improve models, analyse results for given portfolios and display the best solutions for all possible budgets in the form of a Pareto frontier. A case study was performed utilising a system graph and suspected attack paths prepared by industrial security engineers based on an industrial source with which they work.
The goal of the case study is to model a supervisory control and data acquisition (SCADA) industrial system which, due to having a potential to harm people, necessitates strong protection while not allowing to use typical penetration tools like vulnerability scanners. Results are analysed to show how a cyber-security analyst would use \cybopt{}
to store cyber-security intelligence and draw further conclusions.
\end{abstract}



\maketitle

\section{Introduction}

\subsection{Background}

Cyber security needs to be considered by any organisation that uses any kind of IT system, especially when a result of such a system malfunctioning can be tragic -- this is true for most industrial control systems, controlling, for example, power stations and factories. A cyber security \emph{portfolio} is a set of counter-measures against \emph{vulnerabilities} adopted by a company. Such countermeasures are called \emph{controls} in the literature \citep{frameworks, fairytale}. Among others, one of the industry-used frameworks lists these categories of controls: malware defence, controlled use of privileges, penetration tests, and security training programs. \cite{cis_controls}

An IT system can be modelled as an attack graph -- in this particular case study, a \emph{privilege graph} where each node represents a set of privileges, and each edge represents the attacker gaining additional privileges by exploiting a vulnerability. 


\cite{attackgraphs} described a method to obtain min-max optimisation over probabilistic attack graphs, enabling to obtain a portfolio giving the best security (minimised maximum flow to the target) possible for a given budget. Having produced the mathematical theory, the next step is to build an attack graph with probabilistic flows (attack success probability) over edges and controls with defined flow reduction (that is, reduced probability of a successful attack when we employ a security measure).

A crucial application area for cybersecurity decision support is a ClearSCADA-based industrial control system. This ICS is software that runs in factories and controls mechanical devices, so an exploit in the system may pose a danger to the workers and the general community (e.g. if the system manages essential services like energy or water supply). Therefore strong cybersecurity protections are necessitated. A factory system graph and suspected attack paths were received from industrial security engineers based on their work with an industry source. This detailed and trusted information proved crucial to a successful analysis.

\subsection{Contribution}

\begin{figure}
    \hspace*{-0.4cm}\includegraphics[width=1.1\columnwidth]{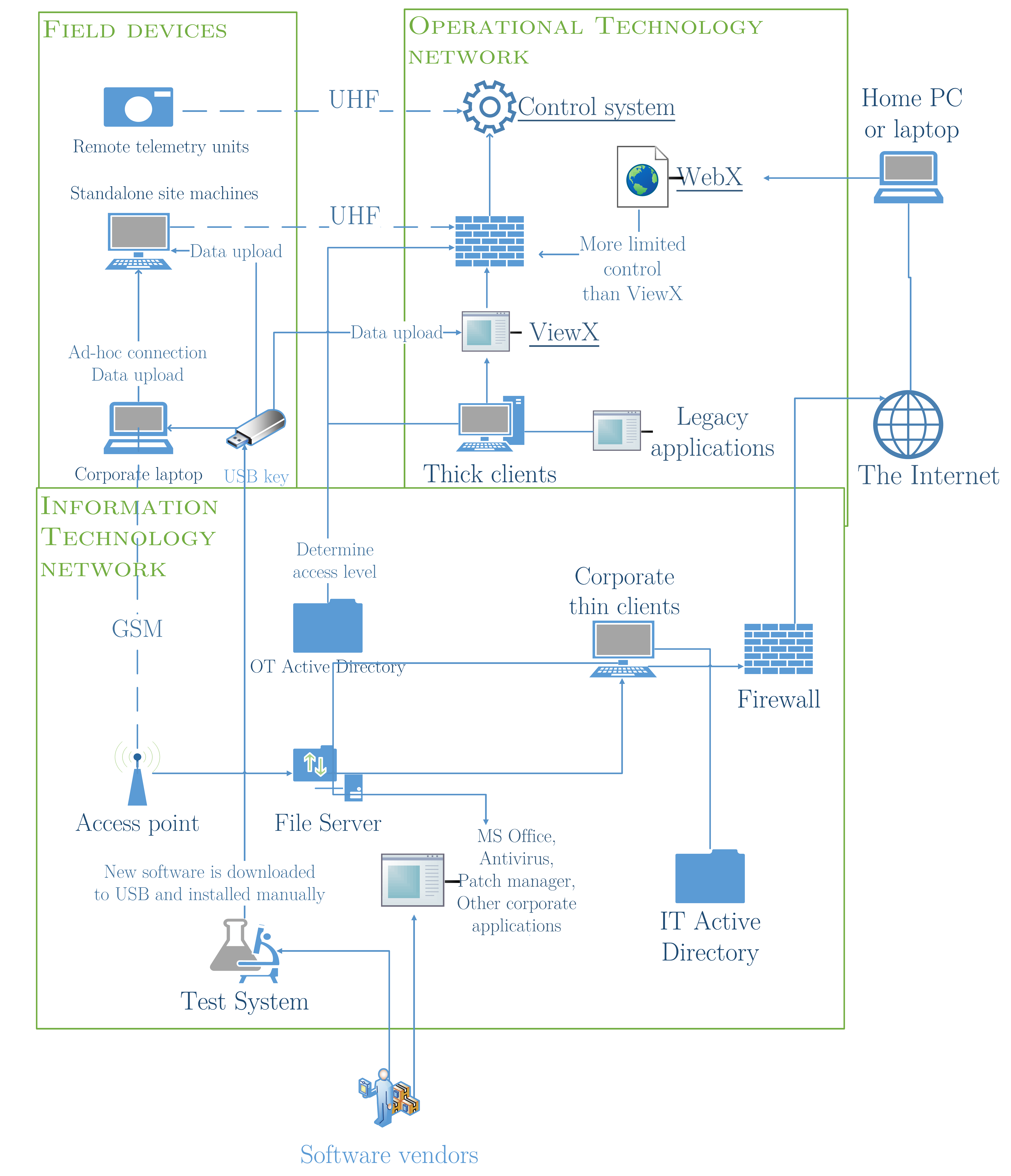}
	\caption{The industrial case study analysed, divided into three segments: OT, IT and field devices. Underlined are the control system and its two main interfaces: WebX and ViewX.}
	\label{fig:diagram_visio}
\end{figure}

This work reports on a tool implementing the min-max optimisation from \cite{attackgraphs}.
Design choices and features of the tool are explained, and an application of the tool to a real-world SCADA system is shown. More specifically, the goals are to:


\begin{itemize}[leftmargin=*]
	\item Build a software package (\cybopt) with the following main features:
		\begin{itemize}[leftmargin=*]
			\item Visualisation of attack graphs;
			\item Creator and editor of attack graphs;
			\item Tight integration with the optimisation algorithm.
		\end{itemize}
	\item Demonstrate the effectiveness of the approach by developing a case study (an overview of the system analysed is shown in Fig. \ref{fig:diagram_visio}):
		\begin{itemize}[leftmargin=*]
			\item Model an attack graph based on the information received;
			\item Assume the role of a security analyser and make some sound conclusions from the model.
		\end{itemize}
\end{itemize}

\subsection{Paper structure}

The paper is split into three main parts: 1) the literature review, discussing the developments in the area of attack graph modelling and security of industrial control systems; 2) documentation of the design choices made while developing \cybopt{}; 3) modelling a SCADA system and the findings made by feeding it to the optimisation routine using \cybopt{}. 


\section{Literature review}

\subsection{Attack graph modelling}

\cite{Protection} is credited with the idea of access control modelling. In this paper, the author models access as an \emph{access matrix}. 
It was found out that the safety problem --- checking if a subject (i.e. user) can get access to an object is insoluble under an access matrix model. \cite{protection_inos}


One of the models developed for solubility of the safety problem is the take-grant protection model. Such a graph can be modified at the run-time to model a user \emph{taking} advantage of their existing permissions to \emph{grant} themselves new ones \citep{theoretical_issues}. 
Soon after the model's conception,
\cite{synthesis} has criticised the ability of the model to accurately represent real-life systems which led to the development of other DAG-based models.

As extensively surveyed by \cite{dag_model}, DAG-based models have been since then broadly used to model real-life attack and defence situations. They have identified more than 30 different approaches, most of which can be broadly categorised as derived from \emph{threat trees}, used early by \cite{attack_tree}, and formal methods based on Bayesian networks, for example, probabilistic attack graphs, which allow reflecting the probabilistic nature of security dangers \citep{bayesian_attack_graphs}.

Choosing between security controls, that is, counter-measures against security threats has been shown to be a multi-objective decision problem. \citep{multi_objective} \cite{attackgraphs} have shown a framework to efficiently solve a multi-objective optimisation problem presented as a probabilistic attack graph combined with a set of controls reducing the probabilities. Incorporating this algorithm into a user-friendly tool is the primary objective of this paper.

\subsection{Modelling tools}

The next step in the popularisation of computer security models is to implement tools allowing to analyse of systems using these formal methods. One of the first programs for this purpose was a C++ attack graph generator for Unix, written by \cite{graph_generation}. This tool took as an input \emph{attack templates} (descriptions of attacks consisting of privileges and capabilities necessary and obtained as a result), \emph{configuration profile} (i.e. network topology, privilege sets in the system) and \emph{attacker profile} (description of attacker's capabilities). The figures included in the paper bear remarkable similarity to the drawings generated by \cybopt{}, excluding the probabilistic aspect.

Other tools followed, primarily focusing on the problem of generation of attack graphs from realistic network diagrams and vulnerability databases exported from already existing tools. One of these is NetSPA, which generates attack graphs from netViz files, NVD vulnerability databases and exported firewall rules \citep{netspa2}. TVA (Topological Vulnerability Analysis), on the other hand, incorporates a network scanner which is used to generate attack graphs \citep{tva}.

Probably the most similar model to the one implemented in\\ \texttt{CySecTool} was found in \cite{mulval2, mulval}. The authors extended the MulVAL tool to include probabilities of each vulnerability being exploited, using vulnerability metrics like Common Vulnerability Scoring System. The author says that the major disadvantage to applying Bayesian networks to this problem is the very slow calculation and optimisation.

Thanks to the theoretical advantages \citep{attackgraphs} and significantly increased processing power of home computers, it is not as infeasible as it was back then. Parallelisation, which was implemented as a part of this work, also greatly helped.

\subsection{Industrial cybersecurity} \label{sec:scada_review}

\emph{Supervisory control and data acquisition} (SCADA) systems are used to monitor and remotely control critical industrial processes, such as gas pipelines, electric power transmission or other production infrastructures. The term was invented in the 1980s, but the first systems were created in the 1960s \citep{cybersecurity_scada}. SCADA systems are a part of wider \emph{Operational Technology} (OT), which is defined as technology that interfaces with the physical world and includes Industrial Control Systems (ICS), SCADA and Distributed Control Systems (DCS) \citep{ot}.


The significant change in the cybersecurity of industrial control systems was the introduction of Internet Protocol-connected systems in the 1990s. The systems built or extended as such became vulnerable not only to specialised attacks but to all the other common Internet threats, for example, automated crawling of servers in a search for open vulnerabilities. Even if these networks are not connected to the Internet, they may remain as vulnerable -- in 2010, the Stuxnet virus was found to spread using USB drives, and once in, it would spread further using the local IP network \citep{ics_threats}.
Stuxnet, allegedly created by the USA and Israel, specifically targeted SCADA systems and was reported to have caused 20\% of centrifuges used in the Iranian nuclear programme to fail \citep{nyc_stuxnet, ieee_stuxnet}.


New challenges appear a lot faster than an expected lifespan of an Industrial Control System, which for each individual component, both software and hardware, is often over ten years, and occasionally over 20 years. \citep{Hahn2016} \cite{evolution} in an industry report quotes data that 64\% of OT organisations struggle with developments in cyber-security; more concerning, 74\% of the organisations surveyed report that they were breached in the last 12 months.

Modelling industrial control systems has been demonstrated to have some interesting caveats absent otherwise. While one may be tempted to run a vulnerability scanner on a system, one must remember that an industrial system malfunction can have outcomes in the real world, as opposed to, for example, a website that can be easily restored from a backup. \cite{duggan_2005} reports several incidents:
\begin{itemize}[leftmargin=*]
	\item A ping sweep was performed on a SCADA system. Suddenly one robotic arm became active and swung 180 degrees.
	\item A penetration test was performed to itemise hosts on a gas company network. The search ventured into the SCADA part of the network, which was unintended, and locked the system for 4 hours, cutting off the gas supply.
	\item A ping sweep caused a system controlling the creation of integrated circuits to hang, destroying \$50,000 worth of wafers.
\end{itemize}

All three tools mentioned (NetSPA, TVA and MulVAL) depend on a vulnerability scanner. As demonstrated by the examples above, a different approach is necessary. In the previous edition of this workshop, a generator of attack graphs based on the IEC 61850 System Configuration Description Language was shown. \cite{iec} Meta Attack Language is used for modelling, which was shown to enable users to develop probabilistic models of attacks against vehicles, for example, passenger cars. \cite{auto} The next step we think is worth taking is to simplify creating and reasoning about graphs of this type --- \cybopt{} is the result.



\section{\cybopt{} design and implementation}

\cybopt{} is a web application developed using Python and the Bokeh presentation library.
It provides a front end for the optimisation engine implementing the algorithm in \cite{attackgraphs}, which is further explained in \autoref{sec:routine}.

Attack graphs are saved to a JSON file in a bespoke schema, storing data about nodes (attackers' access level), edges (vulnerabilities escalating privileges) and controls (counter-measures against vulnerabilities). Graphs can be edited using an included editor, and the application contains several example graphs, including the Industrial Control System analysis. A JSON schema file is provided, so the models can be validated, viewed, edited and created using not only \cybopt{} but also standard tools with support for JSON documents \citep{json_schema}.

\begin{figure*}[ht]

	\centering
	\includegraphics[width=\textwidth]{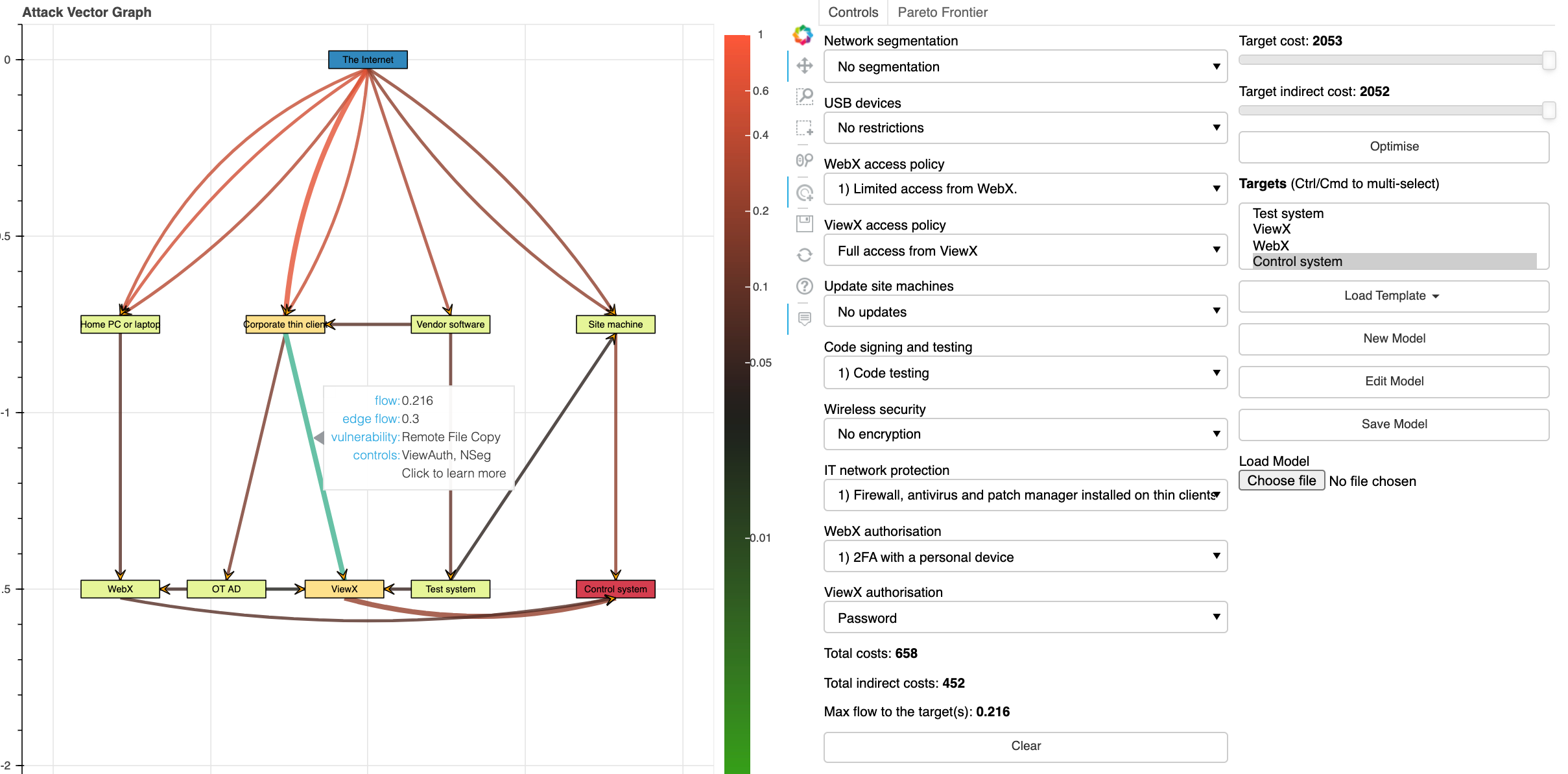}
	\caption{A screenshot of \cybopt{} with the industrial control attack graph. On the right hand side the controls installed.}
	\label{fig:screenshot}
\end{figure*}

A screenshot of \cybopt{} is shown in Figure \ref{fig:screenshot}: on the left-hand side, the attack graph modelling the security scenario under analysis is drawn. Edges are coloured according to the probability of a successful attack, using the logarithmic scale drawn on the right-hand side of the graph. The path drawn in the thicker line is the critical path --- the most likely attack path. The initial attacker's status node is drawn in blue, while their targets are drawn in red. On the right-hand side of the program view, the user can manually select a portfolio, change the targets, load and save a graph, launch a graph editor or optimise for a given budget.

With the click of the ``Edit" button, the active model is loaded to the editor. Once changes are applied, the model can be saved locally or instantly viewed in the visualiser module of \cybopt{}. Due to how fast it is to make changes and observe the results, an analyser can experiment with model parameters very comfortably.

The portfolio can be chosen manually using the panel closely depicted in Figure \ref{fig:screenshot}. Flows in the graph are momentarily recalculated, and one can see what influence the choice they made had -- they can look at the colours, the maximum flow and budget costs displayed on the right-hand side of the graph or hover on the edges in order to see detailed information about the vulnerability represented by the edge hovered on, controls applied, and calculated flows. One can click on an edge to go to a page describing the vulnerability in detail --- in the case of this case study, it leads to the vulnerability description in MITRE ATT\&CK for the ICS database.

The application can generate an optimal security controls portfolio in a specified budget. For a detailed analysis, it lets the user generate a \emph{Pareto frontier} -- the set of all portfolios for which no more secure solution exists in the same or lower budget. Such solutions are called \emph{Pareto-optimal} \citep{pareto}. Charting them lets the user analyse the way security changes with an increasing budget. Hovering on the points lets them get additional data about the portfolio -- its exact cost, indirect cost and security damage (that is, the attack path with maximum probability given the portfolio). Clicking on a point applies the associated portfolio.

\begin{figure}[htbp]
	\centering
	\includegraphics[width=\columnwidth]{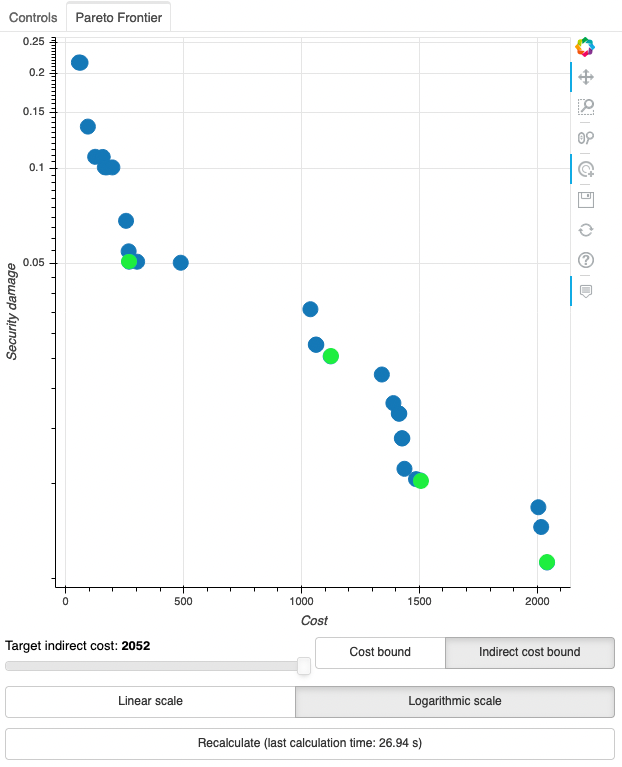}
	\caption{A Pareto frontier of the ``optimal" portfolios applicable to the ICS model. Note the logarithmic security scale and unbound indirect cost. Portfolios in green are further analysed in \autoref{sec:use_cases}.}
	\label{fig:scenario_ics_pareto}
\end{figure}

An example of a generated Pareto frontier can be seen in Figure \ref{fig:scenario_ics_pareto}. Short calculation time --- approx.~10~s on AMD Ryzen 7 PRO 3700U and approx.~18~s on Apple~M1 --- is achieved due to adding parallelisation to the original optimisation routine. Parallelisation was achieved with the help of the pathos framework, while PuLP is used to implement the optimisation routine itself \citep{pathos, pulp}.


Combining automated portfolio optimisation with manual fine-tuning enables detailed analysis and making informed decisions. Graphical representation of the results, generated in real-time, helps to quickly establish the consequences of the choices one can make.

For correctness, a set of tests was prepared using the unittest library and Selenium browser automation tool. Some were based on the manually verified and formally proven results from \cite{attackgraphs}, while others are based on bespoke scenarios created during the development. 

\subsection{Overview of the optimisation routine}
\label{sec:routine}

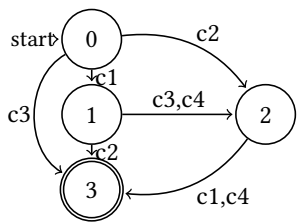
\begin{figure}[ht]
	\centering
	\begin{tikzpicture}[->,=stealth,shorten >=1pt,
		auto,node distance=1cm,semithick,inner sep=1pt,scale=0.1]
		\node[state] (1) {\large 1}; 
		\node[initial,state] (0) [above of=1] {\large 0}; 
		\node[state,accepting] (3) [below of=1] {\large 3}; 
		\node[state] (2) [right=1.5cm of 1] {\large 2}; 
		\tikzstyle{every node}=[font=\large]
		\path (0) edge node {\large c1} (1)
			edge [bend left]  node {\large c2} (2)
			(1) edge node {\large c3,c4} (2)
			(2) edge [bend left] node  {\large c1,c4} (3)
			(1) edge node {\large c2} (3)
			(0) edge [bend right=60] node[left] {\large c3} (3);
	\end{tikzpicture}
	\caption{A toy example of an attack graph. Edges are labelled by the controls which are effective when applied to them.}
	\label{fig:toy example}
    \vspace*{-0.2cm}
\end{figure}

To illustrate the optimisation framework and system modelling implemented in \cybopt, let us consider the toy example in Figure~\ref{fig:toy example}.
The initial state of the attacker is vertex 0, and the target state is vertex 3. An edge in the graph is an attack step or a vulnerability; for example, the attacker could take the leftmost edge and reach the target in one step. Security controls are associated with edges that represent vulnerabilities which they are effective against. For example, for the leftmost edge, control c3 is effective; for the edge from 1 to 2, there are two effective controls: c3 and c4.
Suppose that when deploying c1, the attacker probability of success on that edge is $0.5$, for c2 and c4 it is $0.2$, and for c3 it is $0.1$. 

The optimisation is answering questions like: ``suppose the defender can only choose two security controls out of c1, c2, c3, and c4: which combination is the optimal choice in the sense of making the probability that the attacker successfully reaches vertex 3 the lowest?'' 
 and variations of this question, where controls have varying ``costs" and the defender has a varying ``budget".

For example, if the defender were to choose c3 and c4, then the attacker would reach 3 with probability 1 by following the path (0,1), (1,3). However, if the defender were to choose c2 and c3 instead, then the attacker can reach 3 with a probability 0.2 at most. In fact
c2, c3 is the optimal answer to the question posed above --- the optimal ``portfolio" of security controls.

In mathematical terms, this is a bi-level optimisation: the defender wants to {\em minimise} the attacker's {\em maximal} probability of reaching the target. The minimisation is the {\em outer problem} and the maximisation is the {\em inner problem}. In addition, the defender is subject to budget constraints.
The work \cite{attackgraphs}, using ILP conversion, exact LP relaxation, and dualisation introduces an algorithm to efficiently solve this problem, which is implemented as a backend of \cybopt{}.

\section{Methodology} \label{sec:methodology}

Our attack graphs are defined as follows: 
\begin{enumerate}[leftmargin=*]
	\item vertices (attacker access privileges),
	\item edges (classes of vulnerabilities),
	\item default flows over edges (default probability of the unmitigated vulnerability being exploited), 
	\item controls (mitigations),
	\item flow reductions (reduction of the probability of  the  vulnerability being exploited if control is deployed),
	\item direct costs (monetary cost for implementing a control) and indirect costs (inconvenience to stakeholders, i.e. downtime, increased complexity of use).
\end{enumerate}

For this scenario, all the above items (vertices, edges, etc.) are described in detail in the following 
along with the decisions made and their justifications.

\subsection{Vertices, or the ICS system}

The case study is based on the main operating capacity of a medium-sized utility provider, as described by the national cybersecurity agency. For the purposes of confidentiality, the exact organisation and sector have been obfuscated. 
The example considers a single physical operating environment for the IT and OT (Operational Technology) systems and several remote sites which relate to control units.

For the purpose of documentation, the environment is broken up into four distinct parts, the OT/SCADA system, IT system, Field Devices, and External Devices and Systems. Each of these parts can be defined by a conceptual physical or technological separation, although connectivity does exist between sections --- which will prove vital to the system's security. In this example, all clients and servers are all running operating systems vendored by Microsoft. This may not hold true for the devices external to the network, which may be running other operating systems.



A figure depicting the industrial control system under analysis is shown in \autoref{fig:diagram_visio}. The corresponding attack graph in \cybopt{} is shown in \autoref{fig:screenshot}. 
Most vertices warrant further explaining:

\begin{itemize}[leftmargin=*]
	\item \emph{ViewX} --- one of two front-ends of the system, which is used to control the plant from the site machines and has unlimited capabilities.
	\item \emph{WebX} --- the Internet-facing front-end, which can be used to control the plant from the outside of its perimeter.
	\item \emph{OT Active Directory} --- the part of the directory service which controls access to the control system. Notably, it is controlled from the IT part of the system, so the admins can grant rights to the OT while sitting in the IT.
	\item \emph{Vendor software} --- most of the software, both for IT and OT, gets supplied by external vendors. It is a node of our graph since an attack can occur against the vendor in a ``supply chain compromise" \cite{supply_chain}.
	\item \emph{Test system} --- all the updates to the OT software go through the Test system, which is a miniature of the actual system. Once the software is tested, it gets released to production.
	\item \emph{Site machine} --- machines at the site which communicate with the control system.
	\item \emph{Corprate thin client} --- machines in the IT area, used for administrative tasks, including data preparation for the OT.
\end{itemize}

\subsection{MITRE ATT\&CK for ICS}

MITRE ATT\&CK framework is a threat intelligence database. \citep{getting_started_with_attack, nickels_2018} It collects ``matrices" consisting of:

\begin{enumerate}
	\item tactics (attacker's goals);
	\item techniques and sub-techniques (ways to achieve goals);
	\item mitigations;
	\item procedures (real-life examples of techniques' usage).
\end{enumerate}

MITRE is an organisation that, among others, manages \emph{Common Vulnerabilities and Exposures} and \emph{Common Weakness Enumeration} databases, commonly used in communication about cybersecurity. ATT\&CK is a natural extension of these, creating an overview of tactics and techniques used. While a general, ``enterprise" matrix is by far the best developed, a review process began in 2017 of a matrix for Industrial Control Systems, named \emph{ATT\&CK For ICS}. \citep{alexander_belisle_steele_2020} The matrix is available online for browsing at\\ \url{https://collaborate.mitre.org/attackics}, but the most interesting feature is that the whole database can be downloaded in the form of Structured Threat Information Expression, which is a language and serialisation format based on JSON used to exchange cyber threat intelligence. MITRE provides recipes and tips on how to write programs interacting with that content. \citep{stix}

\subsection{Vulnerabilities}
\label{sec:vulnerabilities}


To analyse the vulnerabilities, we used Jupyter notebooks, a built-in Python JSON processing library and pandas for data processing and display. First of all, we narrowed techniques down to these applicable to tactics modelling gaining access to the control system: initial access and lateral movement. To measure the initial risk, we joined the table with procedures (instances of hacking groups using the vulnerability) and counted how many procedures exist for each vulnerability. Refer to \autoref{tab:procedures} for the list of analysed procedures. The initial risk was calculated as $\frac{|procedures|}{10}$, and it ranges from $10\%$ to $90\%$.

The next step was to add these vulnerabilities to the model, which was done using the built-in \cybopt{} editor. 
Representing multiple attack paths in terms of ATT\&CK allowed us to extract the data about how common they are from ATT\&CK and represent this frequency as the probability of the vulnerability being exploited.

\begin{table}
\centering
\caption{Relevant techniques extracted from ATT\&CK for ICS database.}
\label{tab:procedures}
\begin{tabularx}{\columnwidth}{Xllr}
\toprule
                             technique &                 tactic &                                                                           url &  c.\footnotemark[1] \\
\midrule
           Spearphishing Attachment &     initial access & \href{https://collaborate.mitre.org/attackics/index.php/Technique/T865}{T865} &                 9 \\
           External Remote Services &     initial access & \href{https://collaborate.mitre.org/attackics/index.php/Technique/T822}{T822} &                 8 \\
                Drive-by Compromise &     initial access & \href{https://collaborate.mitre.org/attackics/index.php/Technique/T817}{T817} &                 6 \\
            Supply Chain Compromise &     initial access & \href{https://collaborate.mitre.org/attackics/index.php/Technique/T862}{T862} &                 3 \\
 Engineering Workstation Compromise &     initial access & \href{https://collaborate.mitre.org/attackics/index.php/Technique/T818}{T818} &                 2 \\
Replication Through Removable Media &     initial access & \href{https://collaborate.mitre.org/attackics/index.php/Technique/T847}{T847} &                 2 \\
          Data Historian Compromise &     initial access & \href{https://collaborate.mitre.org/attackics/index.php/Technique/T810}{T810} &                 1 \\
         Internet Accessible Device &     initial access & \href{https://collaborate.mitre.org/attackics/index.php/Technique/T883}{T883} &                 1 \\
           External Remote Services &   lateral movement & \href{https://collaborate.mitre.org/attackics/index.php/Technique/T822}{T822} &                 8 \\
                     Valid Accounts &   lateral movement & \href{https://collaborate.mitre.org/attackics/index.php/Technique/T859}{T859} &                 8 \\
    Exploitation of Remote Services &   lateral movement & \href{https://collaborate.mitre.org/attackics/index.php/Technique/T866}{T866} &                 3 \\
                   Remote File Copy &   lateral movement & \href{https://collaborate.mitre.org/attackics/index.php/Technique/T867}{T867} &                 3 \\
         Program Organization Units &   lateral movement & \href{https://collaborate.mitre.org/attackics/index.php/Technique/T844}{T844} &                 2 \\
\bottomrule
\end{tabularx}
	\footnotemark[1]{\footnotesize Procedures count.}
\end{table}


\subsection{Controls}
\label{sec:controls}

Control is a measure against vulnerabilities. In \cybopt{} it is represented with its name, direct cost, indirect cost and the flow reduction for the vulnerabilities it is effective against. Moreover, controls exist in mutually exclusive levels -- for example, one could opt for updating air-gapped devices' software once a year or once a month.

In order to obtain the controls for each vulnerability, we listed mitigations that are applicable according to the ATT\&CK database and grouped them in the following control groups. Some mitigations already existed in the system; they are underlined. 

\begin{enumerate}[leftmargin=*]
	\item \emph{Network Segmentation}:
		The current network, while segmented in theory into the Internet, OT, IT and site devices, has multiple points of contact between the zones, which allow attacks to happen. Most egregiously, the Active Directory of OT users is in the IT zone, so the admins can easily modify users without having to go to the site. \\
		Data files are also uploaded from the IT to the OT, effectively controlling the control system.
		Levels: 1)
			\underline{No segmentation};
			2) ViewX and OT Active Directory removed from IT zone;
			3) Access to OT from IT limited to WebX.
	\item \emph{IT Network Protection}:
		Thin clients are used in the IT zone for administrative purposes. Since a connection exists between IT and OT, an infected IT computer could wreak havoc in the ICS. Therefore, we want to minimise the risk of this happening. All of these protections have been applied in IT.
		Levels: 1) No protection; 2) \underline{Firewall and antivirus installed on thin clients.}
	\item \emph{USB Devices}:
		Currently, USB devices are used for the manual carrying of data to and from OT. While it allows us to air-gap devices, the system is still open to Stuxnet-like lateral movement. Levels: 1)
			\underline{No restrictions}; 2) Company issued and periodically formatted drives only; 3) Single-use CDs only recorded directly by SCADA software.
	\item \emph{WebX Access Policy}:
		WebX is used to allow limited control from outside of the site. It reduces the need to go to the site to do some management, but it opens the system to attacks from the Internet. 
			Levels: 1) Full access; 2) \underline{Limited access}; 3) WebX reduced to site monitoring.
	\item \emph{ViewX Access Policy}:
		ViewX is available only from the site but can do everything, and there are no users. Creating more access levels means that when access credentials are leaked, it does not always mean that the attacker gains full control. Levels: 1)
		\underline{Full access}; 2) Access levels introduced.
	\item \emph{Update site machines}:
		Machines without an Internet connection are very rarely updated. While they cannot get infected from the Internet, a virus can well get there from a USB drive. Distributing updates would require a person to update all these machines manually. Levels: 1)
		\underline{No updates}; 2) Yearly updates; 3) Monthly updates.
	\item \emph{Code signing and testing}:
		Enforcing that only valid and tested code can run on the machines can foil many possible attacks but adds a considerable overhead for deploying new code. Moreover, machines can be too old to support only signed binaries --- it was introduced with Windows 7/2008 R2 \citep{applocker}. We could update these, but updating PLCs would require replacing the factory equipment. Levels: 1)
			No code testing; 2)
			\underline{Testing all updates}; 3)
			Testing and code signing on PCs; 4)
			Code signing on PCs and PLCs.
	\item \emph{Wireless security}:
		Since the communication between site machines and the OT is going over unencrypted UHF radio waves, our adversary could catch these communications, modify and replay them to bring the site out of control. A Polish student did that, using a modified TV remote controller to control a tram system in Łódź. \citep{rmf-fm}
		Encrypting network traffic is noted to be expensive since the radio devices have no ability to handle it. DES could be used without having to replace the whole radio infrastructure. Levels: 1)
			\underline{No encryption}; 2)
			DES encryption; 3)
			AES encryption.
	\item \emph{WebX authentication}:
		Currently WebX can be accessed from any external device what is inherently insecure. The factory could provide 2FA keys or separate devices to the employees which give them access only to WebX. Levels: 1)
			Username/password; 2)
			\underline{2FA with a personal device}; 3)
			2FA with a physical key;
		4)	Company-provided devices used.

	\item \emph{ViewX authentication}:
		Currently ViewX is protected only by a single password. There are no accounts. \\
			Levels: 1) \underline{Password};
			2) 2FA with a personal device;
			3) 2FA with a physical key.
\end{enumerate}

\subsection{Controls' cost}

In order for the optimiser to find the best solution in a budget, we need to assign to all controls their associated cost. Moreover, the cost does not have to be only monetary --- if we make it harder for the employees to do their jobs by, for example, taking away their access rights, there is a loss of productivity, which can be modelled using \emph{indirect cost}.

Many controls can be divided roughly into ``direct cost-intensive" and ``indirect cost-intensive". For example, installing new radio devices, which would allow us to encrypt their communication, would be very expensive, but the employees should not even notice it. On the other hand, restricting the capabilities of WebX (the Internet-facing front-end to the control system) is very cheap to do, but the employees will be greatly annoyed by having to go to the site to do their jobs, especially in 2021 when they are told to work from home. The system is very flexible, so the analyser can easily adapt the costs to the ever-changing environment.

We express mitigations in some basic actions, like ``Employing extra employee" or ``New infrastructure". This way, the costs of the basic blocks can be adjusted, and it is easier to compare the mitigations. The actions used are listed in \autoref{tab:actions}.

	\begin{table}
		\caption{Basic actions, their direct, indirect costs and rationale behind them.}
		
		\begin{tabularx}{\columnwidth}{p{1.5cm}llX}
			\toprule
			Action & D.\footnotemark[1] & I.\footnotemark[2] & Rationale \\
			\midrule
			Buying a tool, e.g., USB sticks   or a laptop. & 1   & 8  & Buying 50 pen drives is   comparatively cheap. Dealing with employees’ complaints about the new   procedures is not.                                                                                                               \\
			Extra employee.                                & 5  & 0  & Extra person to handle the work  due to a new policy. Notably, an extra administrator  to work at the site due to the limited external access policy falls into this   category. \\
			New PC software.                               & 2   & 8  & The cost of a new subscription   is more painful than buying some USB sticks. The employees’ have to deal with the change.                                                                 \\
			New ICS software feature.                      & 40  & 20  & A new feature in a bespoke ICS software is going to be more expensive to develop than buying an off-shelf MS   Office box.                                                                                                       \\
			New PLC infrastructure.                        & 100 & 0  & What we want to avoid the most   is changing elements of the bespoke PLC infrastructure. Such changes remain   transparent to the employees – hence the negligible indirect cost.                                                  \\
			Taking away permissions.                       & 1   & 40 & Taking away permissions from   existing users reduces their productivity. \\
			\bottomrule
		\end{tabularx}
		\label{tab:actions}

		\footnotemark[1]{\footnotesize (Direct) cost.}
		\footnotemark[2]{\footnotesize Indirect cost.}
	\end{table}


\subsection{Controls' effectiveness}

The controls' effectiveness is the part of the model in which, for each vulnerability and control, a number in $[0; 1]$ is given, specifying the effectiveness of the control against the vulnerability. This number is named \emph{flow reduction} and is the number by which the initial flow on the edge (see \autoref{sec:vulnerabilities}) is multiplied when the control is active.

Intuitively, flow reduction equal to one means that the control does not impact the risk; equal to zero means that it eliminates the risk completely; and equal to, for example, 0.5 halves the risk.


We graded the efficacy of each of the controls on a 4-point scale (low, medium, high, very high), associated flow reduction with each point (0.7, 0.5, 0.2, 0.1).

To show some examples:
\begin{itemize}[leftmargin=*]
	\item \emph{DES Encryption} -- medium, \emph{AES Encryption} -- very high. While everything is better than plain-text, once the attacker finds out DES is used, they can use an online GPU cluster to break it, while AES encryption is considered secure \citep{symmetric_encryption} -- so a~compromise would have to occur due to an implementation error.
	\item \emph{Yearly updates} -- low, \emph{Monthly updates} -- medium. While updating the systems protects against common threats, ICS attacks are usually made to order, decreasing the probability that their signature would be detected. Therefore, the updates' protection is on the lower end of the scale.
\end{itemize}


The effectiveness, in the end, corresponds to the opinions of the analysts. It is planned to verify the model with the agency specialists to improve the accuracy of the model further. The proposed way to work with the system is to continuously refine the model and its probabilities in order to reflect our improving understanding of the system and verify it using the Pareto frontier. The continuous improvement approach is shared with the ATT\&CK database. \citep{mitre_philosophy, alexander_belisle_steele_2020}.

Alternatively, once an appropriate data source is obtained, a statistical approach, similar to the one in \autoref{sec:vulnerabilities}, can be followed.

\section{Use cases and results}
\label{sec:use_cases}

Given the security modelling described so far, we are going to show some use cases of \cybopt\, which are of interest to security analysts and decision-makers.

The following scenarios could be relevant to a security analyst using \cybopt\ to generate portfolios to be shown to the stakeholders --- the protections they are offering and their weak points.

In these scenarios, we consider the Pareto frontier generated by the tool and will look at some interesting portfolios in the frontier, explaining how decisions can be made.

\subsection{General analysis and findings}
\label{sec:general_analysis}

Let us consider the frontier in \autoref{fig:scenario_ics_pareto}. Most crucially, it displays how with the increasing budget, we can decrease the security damage, that is, the probability of the most probable attack.

There are several intuitive rules an analyser may follow to narrow down the portfolios to the most interesting ones:

\begin{itemize}[leftmargin=*]
	\item If points are close on the x-axis (cost) but distant on the y-axis (security), a small investment improves the security considerably. Therefore, the point further to the right will be \emph{more} interesting.
	\item Conversely, if the points are distant on the x-axis but close on the y-axis, a significant investment improves the security slightly. Therefore, the point further to the right will be \emph{less} interesting.
	\item Then, if the points are close on both axes, they are similar in both cost and security. These can also warrant further analysis since the control set used can be different, and so the indirect cost, and so the final decision on which portfolio to employ.
\end{itemize}

Using the first two rules, we can find several portfolios representative for ``low", ``medium", ``high" and ``very high" costs. Portfolios, chosen using this method are marked in green in \autoref{fig:scenario_ics_pareto} and detailed in \autoref{tab:general_analysis_tab}.

\begin{table}[!htbp]
	\caption{Portfolios representative for cost categories generated on a naked model.}
	\label{tab:general_analysis_tab}
	\begin{tabularx}{\columnwidth}{l|X|l|l|l}
		\toprule
		Name & Controls & D.\footnotemark[1] & I.\footnotemark[2] & Damage \\
		\midrule
		Naked & No controls & 0 & 0 & 0.27 \\
		Low & Company issued drives only, WebX reduced to site monitoring, ViewX access levels, Wireless DES encryption, ViewX 2FA with a physical key & 269 & 898 & 0.05 \\
		Medium & Code testing, Wireless AES encryption, Company issued drives only, WebX reduced to site monitoring, ViewX access levels, Update site machines monthly, IT network protection, WebX 2FA with a personal device, ViewX 2FA with a physical key & 1124 & 1230 & 0.025 \\
		High & Network segmentation, Single-use CDs only, WebX reduced to site monitoring, ViewX access levels, Code signing on PCs, Wireless AES encryption, IT network protection, WebX 2FA with a physical key, ViewX 2FA with a physical key & 1505 & 1868 & 0.01 \\
		Full & All controls employed at their highest levels & 2042 & 2028 & 0.005 \\
		\bottomrule
	\end{tabularx}
	\footnotemark[1]{\footnotesize (Direct) cost.}
	\footnotemark[2]{\footnotesize Indirect cost.}
\end{table}

\subsubsection{Low-cost portfolio}
\label{sec:low_cost}

The low-cost portfolio is notable for its high indirect costs. Internet access to the control system is reduced to monitoring only, access levels are introduced to ViewX, and ViewX users are required to use physical keys for two-factor authorisation. All of these impact employees' productivity --- to do their job, they must be at the site, ask to have appropriate permissions granted, and pick up a key. If they want to upload data to air-gapped computers, they cannot use their own USB drives anymore. None of these had to be done when there were no controls --- they opened their laptop from wherever and logged in with their password. They could have prepared the data at home to carry it straight to the factory.

Nevertheless, all these controls are very effective-- attacks from the Internet are the most concerning due to how frequent and generic they are, but damage can be limited to information leakage by severely restricting WebX capabilities. Even when malware infects the site computers, it can try to guess users' passwords, but bypassing 2FA is harder.
Moreover, they are very cheap to implement, so if the company does not want to enact controls requiring expensive infrastructure changes, it may be a good solution.

\subsubsection{Medium-cost portfolio}

In this portfolio, due to the higher budget, some high-value investments were made. Notably, AES encryption is added (requiring a replacement of the radio infrastructure), and a test system needs to be built, so new software can be tested, limiting an impact of a supply chain compromise.

The direct cost is now at a level similar to the indirect cost, indicating that these changes require a considerable infrastructure budget and that controls implementable as policies have been largely exhausted. Note that, despite roughly five times higher direct cost, the damage is only two times lower than in the low-cost portfolio.

\subsection{Analysing similar points}

Let us come back to the low-cost portfolio described in \autoref{sec:low_cost}. The main problem with this portfolio is the very high indirect cost in the form of making interacting with the system much more time consuming for the employees.

\begin{figure}[htbp]
	\centering
	\includegraphics[width=0.3\textwidth]{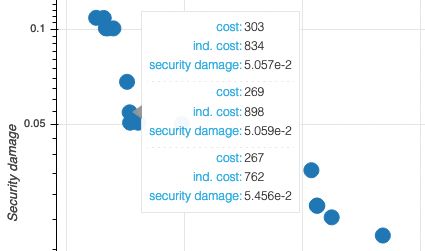}
	\caption{The close-up of the \emph{neighbourhood} of the low-cost portfolio chosen in \autoref{sec:low_cost}. Tooltips give detailed information about the costs and security damage of individual portfolios.}
	\label{fig:portfolio_neighbourhood}
	\vspace{-4mm}
\end{figure}

One of the ways to solve this problem using \cybopt{} is to have a look at the portfolios in a similar price and security range. Such a close-up on three portfolios is illustrated by a screenshot from \cybopt{} in \autoref{fig:portfolio_neighbourhood}.

The initially chosen portfolio is the middle one of the portfolios detailed in \autoref{fig:portfolio_neighbourhood}. Both other portfolios have a lower indirect cost. Let us list the differences in the controls used in \autoref{tab:alternative_portfolio}.

\begin{table}[H]
	\caption{Alternative controls in the neighbourhood of the low-cost portfolio.}
	\begin{tabularx}{\columnwidth}{r|X|r|r|r}
		\toprule
		No. & Differing controls & D. & I. & Damage \\
		\midrule
		1. & Company issued USB devices only, 2FA with a physical key for ViewX. & 269 & 898 & 0.05059 \\
		2. & Company issued USB devices only, 2FA with a personal device for ViewX, IT network protection. & 267 & 762 & 0.05456 \\
		3. & Monthly updates to site machines, 2FA with a physical key for ViewX. & 303 & 834 & 0.05057 \\
		\bottomrule
	\end{tabularx}
	\label{tab:alternative_portfolio}
\end{table}

From the Pareto frontier, it is clear we can go two ways from the first, originally selected portfolio. The second portfolio has a considerably lower indirect cost, similar direct cost and slightly lower security. In contrast, the third portfolio has a higher direct cost, lower indirect cost and comparable security to the first one. Similar to the old idea of the project management triangle, in which one has to choose two from ``good, fast, and cheap" \citep{triangle}, the three portfolios represent adding indirect cost, sacrificing security and adding direct cost.

The clusters of points on the Pareto frontier are helpful because they help to identify which controls, given the model, give similar security. 
In this particular example, we can see that the following pairs of sets are considered comparable in terms of provided security:
\begin{itemize}[leftmargin=*]
	\item \{ViewX physical key 2FA\} and \{ViewX personal device 2FA, IT network protection\}. This reflects the two points of protection -- in the first set, the control system is strongly protected, while in the second, some of that is sacrificed to protect the IT network, from which OT can be infected. It is easier to have the employees use their own devices for 2FA, hence the lower indirect cost.
	\item \{Monthly updates to site machines\} and \{Company issued USB devices only\}. This reflects the problem with malware spreading through USB devices. One way is to restrict USB devices usage, and another is to update even the air-gapped systems regularly, so they are not vulnerable.
\end{itemize}

\subsection{Naked and clothed model}

It is possible that there already exist some controls in the analysed system, while the Pareto frontier assumes there are no controls. In \autoref{sec:controls}, already applied controls are typed in bold.

In the early parts of the analysis, the applied controls were assumed to be in the system and all controls to be built upon them. 
We now remove these protections, creating a so-called \emph{naked model}, so a comparison can be drawn between what the optimiser suggests and the protections already in the system.

\begin{figure}[htb]
 	\centering
 	\includegraphics[width=\columnwidth]{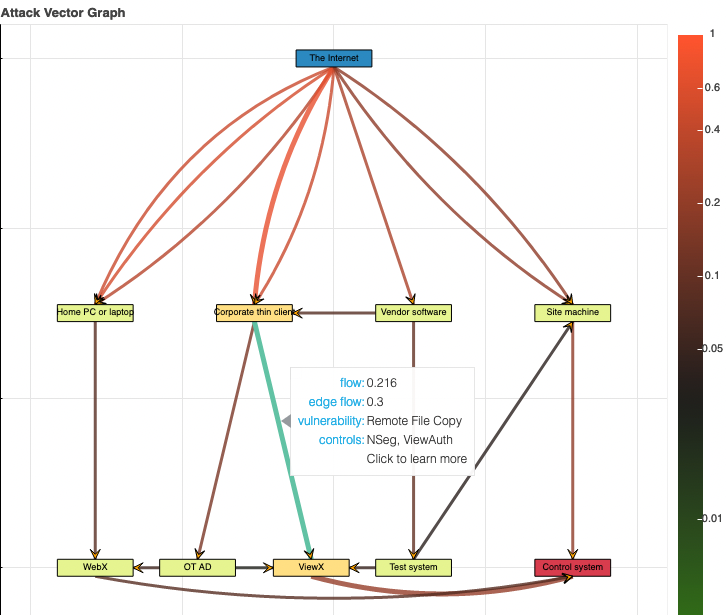}
    \caption{A portfolio chosen to match the real-life system on which the case study was done. A vulnerability on the critical path is hovered on. Refer to \autoref{sec:controls} to see the controls chosen.}
 	\label{fig:scenario_clothed_ics}
 \end{figure}

In the \autoref{fig:scenario_clothed_ics}, a portfolio matching controls already applied in the real system were selected. When compared with portfolios generated in \autoref{sec:general_analysis}, it is clear that, in the model, the current controls are neither secure nor cost-effective.

In the analysis of such cases, it is helpful that \cybopt{} marks the critical path, that is, the attack path with the highest probability, with a thicker line. One vulnerability, Remote File Copy, was hovered on to display detailed data. When clicked on, we are taken to the source database, where we can learn that attacks of this kind on Industrial Control Systems were first observed in 2017, under the names WannaCry, Bad Rabbit, NotPetya. \citep{remotefilecopy, ics-implications}

Since Industrial Control Systems have lifespans counted in decades, many of the vulnerabilities and security practices did not exist when the system was first designed. Indeed, two controls that could help with the vulnerability in question are listed as a challenge faced by ICS systems: poor network segmentation and limited access control. These challenges are said to happen due to an evolution of the cyber threats landscape in Operational Technology \citep{evolution}.

This is where \cybopt{} may shine. With new threats discovered, the basic structure of the model can remain the same, but the probabilities of attacks may change -- and new ones can be added. Therefore, cyber security intelligence can be stored in \cybopt{} models and updated on demand. This is the goal that is shared with the ATT\&CK database, which is intended to be a living project, adapting to the changing conditions \citep{mitre_philosophy, alexander_belisle_steele_2020}.

\section{Conclusion}

The primary goal of this work was to demonstrate the feasibility of an optimisation approach to the problem -- choosing a security portfolio for a ClearSCADA system -- while suggesting other areas where it may be useful.
From a cyber-security point of view, \cybopt{} may be useful to answer questions like -- how many network systems of a given type are vulnerable to common attacks? What is the impact of a newly-discovered vulnerability? What kind of redundancy would be needed to provide security against it? How has network security changed over the years, and what are the dangers of the years to come?

A practical tool based on the optimisation routine was developed. Such a tool may be used by security professionals to demonstrate to their clients not only the kind of services they provide but also explain the way vulnerabilities, both technical and organisational, can be avoided, and if they are not, what would be their impact. Since the model can be adjusted to a particular type of service, it would make the decisions and the budget approved seem more justified to the management, improving overall security awareness.

For researchers and security practitioners, it will allow for a faster information flow from the "academic" to the "practical" side and vice versa. The academics would be enabled to analyse in detail the newest reports, as we have done, and pass the results back. 

In future research, we hope to see more models applied to the different areas -- both broad, such as general ICS analyses, and narrow, for example, modelling a newly discovered web browser vulnerability. What needs further attention is the validation of the results. Decisions made by the system should be compared against decisions made by practitioners, and the models should be refined in practice. Overall, our core finding is that \cybopt{} is ready to be tested in real conditions and have a positive impact on the security of not only Industrial Control Systems but also of other networks.

\begin{acks}
This work was partially supported by \grantsponsor{EPSRC}{Engineering and Physical Sciences Research Council}{https://epsrc.ukri.org/} grants \grantnum{EPSRC}{EP/R002983/1} and \grantnum{EPSRC}{EP/R004897/1}.
\end{acks}

\bibliographystyle{IEEEtranN}
\bibliography{main.bib} 





\end{document}